\documentclass[
    ,final            
  ]
  {aipproc}

\layoutstyle{6x9}

\usepackage{epsfig}

\begin{document}

\title{Mechanical Model for Relativistic Blast Waves and Stratified Fireballs}




\author{Zuhngwhi Uhm}{
  address={Physics Department and Columbia Astrophysics Laboratory, 
           Columbia University, 538 West 120th Street, New York, NY 10027. 
           Email: zu@astro.columbia.edu, amb@phys.columbia.edu}
}

\author{Andrei M. Beloborodov}{
  address={Physics Department and Columbia Astrophysics Laboratory, 
           Columbia University, 538 West 120th Street, New York, NY 10027. 
           Email: zu@astro.columbia.edu, amb@phys.columbia.edu}
}

\begin{abstract}
We propose a simple mechanical model for relativistic explosions with both 
forward and reverse shocks, which allows one to do fast calculations of GRB 
afterglow. The blast wave in the model is governed by pressures $P_F$ and 
$P_R$ at the forward and reverse shocks. We show that the simplest assumption 
$P_F=P_R$ is in general inconsistent with energy conservation law. The model 
is applied to GRBs with non-uniform ejecta. Such "stratified fireballs" are 
likely to emerge with a monotonic velocity profile after an internal-shock 
stage. We calculate the early afterglow emission expected from stratified 
fireballs.
\end{abstract}

\maketitle


\section{Introduction}

Relativistic blast waves from GRBs are believed to produce the observed
afterglow emission of the bursts. The blast-wave structure is 
schematically shown in Figure~1. 
To a reasonably good approximation, the whole blast [the region between 
FS and RS, including the contact discontinuity (CD)] may be assumed to have 
a common Lorentz factor $\gamma_s$ \cite{Piran}. 
The Lorentz factors of the forward shock (FS) and reverse shock 
(RS) are denoted by $\gamma_F$ and $\gamma_R$, respectively. 
Pressures $P_F$ and $P_R$ behind the forward and reverse shocks are calculated 
from the jump conditions.


\begin{figure}[h] 
   \epsfig{file=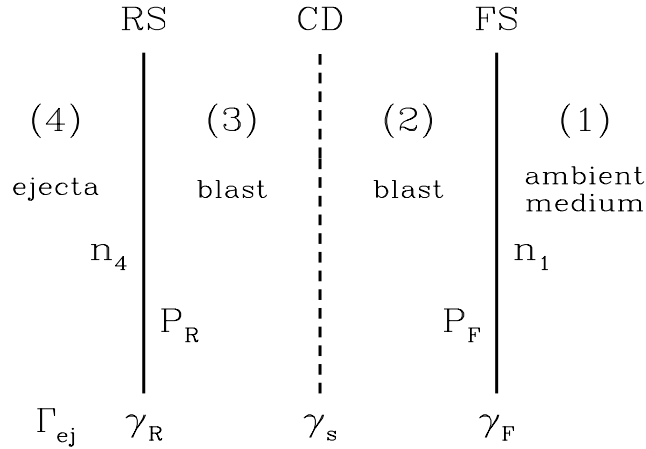, height=14pc, width=20pc}
   \caption{An illustrative diagram of 4 regions in a blast wave.
Here $n_1$ and $n_4$ are the number densities of the ambient medium 
and the ejecta, respectively.} 
\end{figure}


\section{Stratified fireball}

We focus here on the early stages of the GRB explosion when both
forward and reverse shocks are expected to produce significant 
emission. Our calculations assume spherical symmetry. 
However, the results also apply to beamed explosions at early stages,
when $\gamma_s>\theta_{jet}^{-1}$ where $\theta_{jet}$ is the opening 
angle of the ejecta. 

The reverse shock (RS) is expected to propagate in the ejecta and produce 
significant emission at radii $r\sim 10^{16}-10^{17}$~cm. 
The RS emission depends on the structure of the ejecta that forms at
smaller radii $r<10^{16}$~cm, when internal shocks take place \cite{Piran}.

 
\begin{figure}[h] \centering $\begin{array}{cc} 
   \epsfig{file=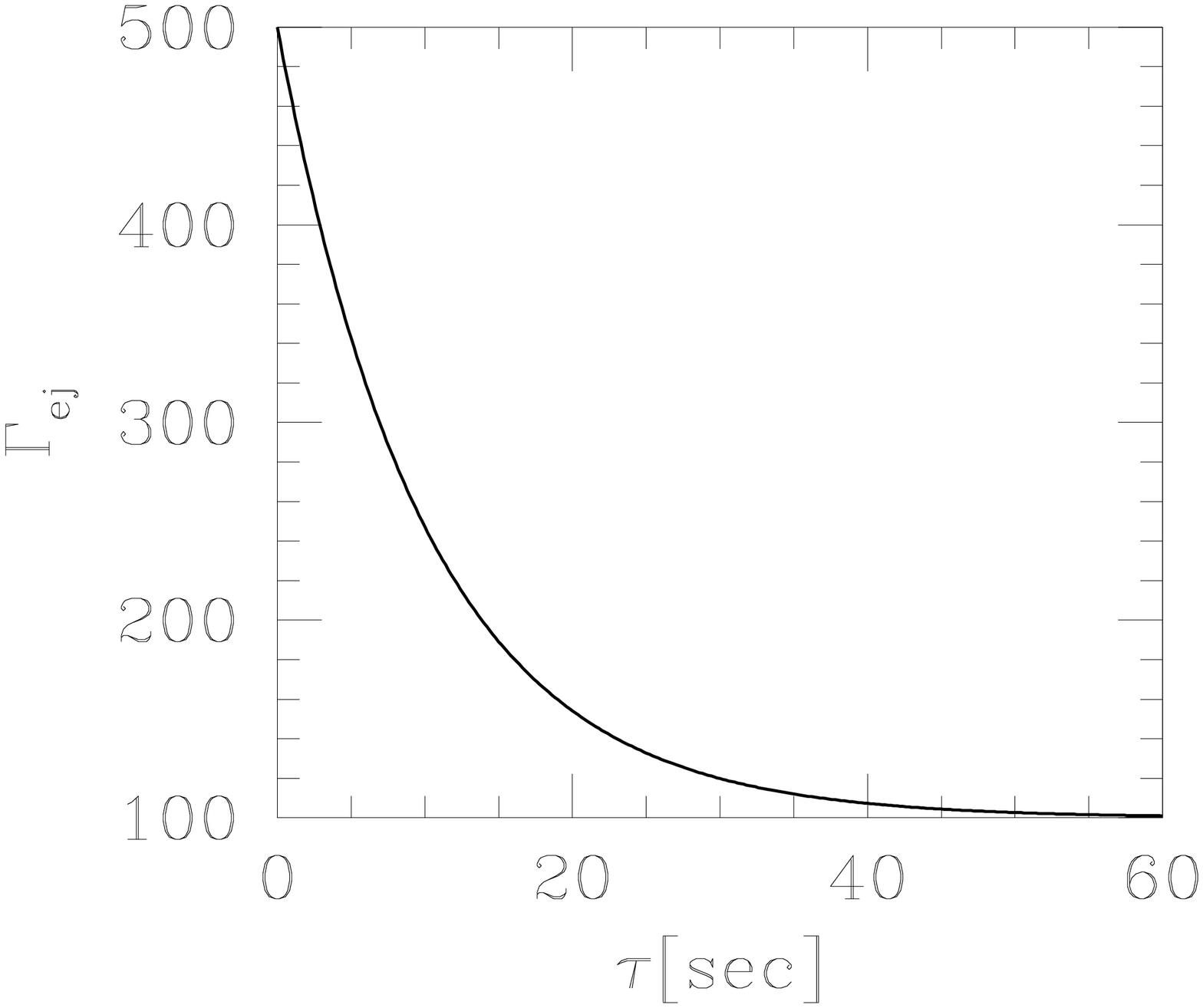, height=11pc, width=14pc}~ &
   \epsfig{file=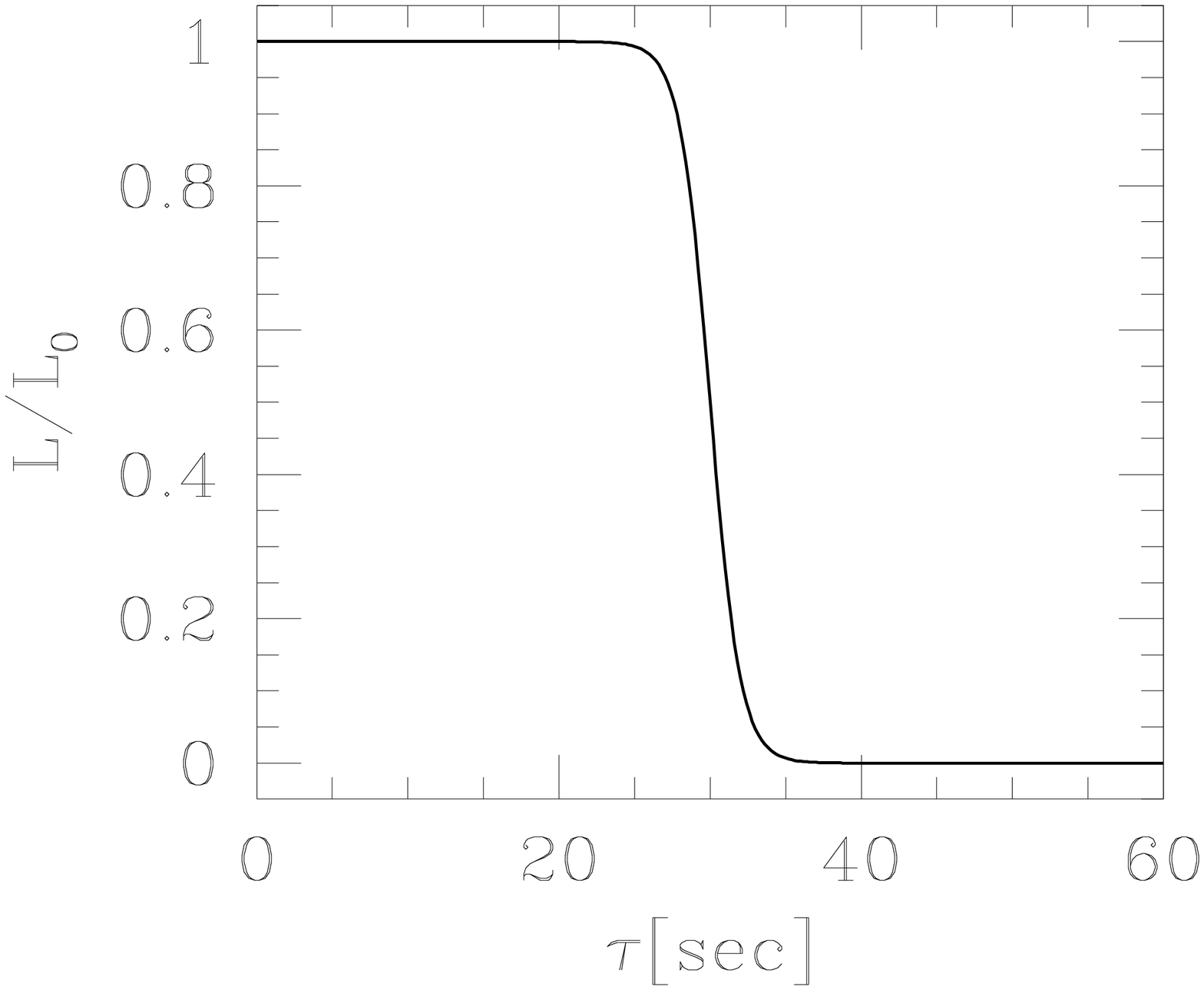, height=11pc, width=14pc} \end{array}$ 
   \caption{
Lorentz factor ({\bf Left}) and kinetic luminosity ({\bf Right}) of ejecta
}      
\end{figure} 

Theoretically, one expects the ejecta to  emerge with a monotonic
velocity profile (internal shocks tend to make the profile monotonic).
Therefore we study here radially stratified ejecta and calculate the
emission expected from such ``stratified fireballs.''

A simple example model considered here assumes 
ejecta with radial profile of Lorentz factor
$\Gamma_{ej}(\tau) = (\gamma_1 - \gamma_2) e^{-a \tau/\tau_b} +\gamma_2$
shown in Figure 2
($\gamma_1 = 500 $, $\gamma_2 = 100$, $a = 3$, and $\tau_b = 30$~s 
is the burst duration).  
Here $\tau$ is the Lagrangian coordinate that
labels the shells in the ejecta by their ejection times.
The luminosity of the ejecta 
$L(\tau)=4\pi r^2 ~ v_{ej} ~ \Gamma_{ej}^2 ~ n_4 ~ m_p ~ c^2$ is
also shown in Figure 2. Specifically,
  $L(\tau) = L_0/(1+\exp[(\tau - \tau_b)/\tau_d ])$ is assumed here 
with
  $L_0 = 10^{52}$~erg/s and $\tau_d = 1$~s for the timescale  of decay.
The total energy of the burst is given by 
$E_{\rm ej}=L_0 \tau_d \ln(1 + e^{\tau_b/\tau_d})$.


\section {Modeling the Blast Wave}

\subsection {1. Customary approximation: $P_F = P_R$} 
              
The customary approximate model blast wave assumes equal
pressures at the forward and reverse shocks: $P_F=P_R$ \cite{Piran}. 
This assumption, as we show in an accompanying paper, enables a simple 
analytical solution for $\gamma_s$ and pressure $P=P_F=P_R$:

\[ \gamma_s = \Gamma_{ej} \left[ (1 - 2\sigma) + 2 \left(\sigma^2 
  + (\Gamma^2_{ej} -1) \sigma  \right)^{1/2}  \right]^{-1/2}  
 \qquad \mbox{where} \qquad \sigma=\frac{n_1}{n_4},
\] 
\[ P=\frac{4}{3}(\gamma^2_{s}-1)n_1 m_p c^2,   \] 

It turns out however that the assumption of equal pressure leads to
a contradiction with the law of energy conservation. The total energy of 
the blast wave and ejecta at a time $t$ may be found by integrating the
$00$ component of the stress-energy tensor over volume, $E = \int T_{00} dV$.
The result is shown in Figure 3.
$E$ decreases with time (dashed curve in the left panel), i.e. 
the energy is not conserved.
We therefore propose another dynamical description.


\subsection {2. Consistent mechanical model}

Instead of finding $\gamma_s$ from $P_F=P_R$ we solve the differential 
equation for $\gamma_s$:
\[ \frac{d \gamma_s}{d t} = \frac{1}{\gamma_s H} v_s (P_R-P_F)
                            -\frac{1}{H} (\gamma_s - 1/\gamma_s) \int^{FS}_{RS} \left(\frac{dP}{dt}\right)dr, \]                                                                       
where $H = \int (U+P) dr$ is the integrated enthalpy of the blast.
The first term on the right describes the external force applied to the
blast due to the difference between $P_F$ and $P_R$, and the second term
describes the adiabatic acceleration of the blast. This equation is 
derived from the exact relativistic fluid equation. It is obtained by 
integrating over the blast and may be viewed as a mechanical equation.
The mechanical model is consistent with the energy and momentum conservation 
(solid curve in the left panel of Fig. 3).

 
\begin{figure}[h] \centering $\begin{array}{cc} 
   \epsfig{file=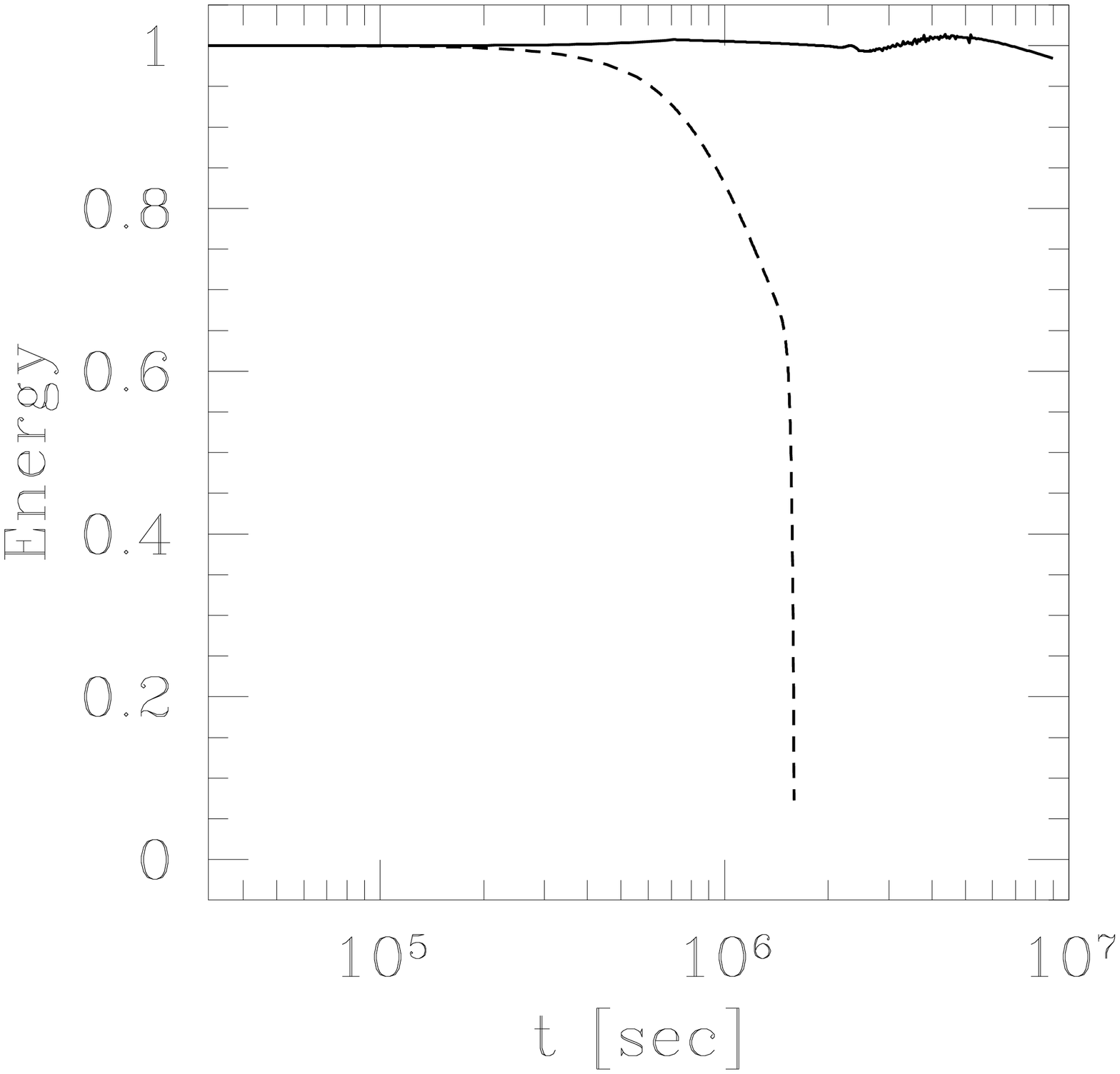, height=12pc, width=15pc}~ &
   \epsfig{file=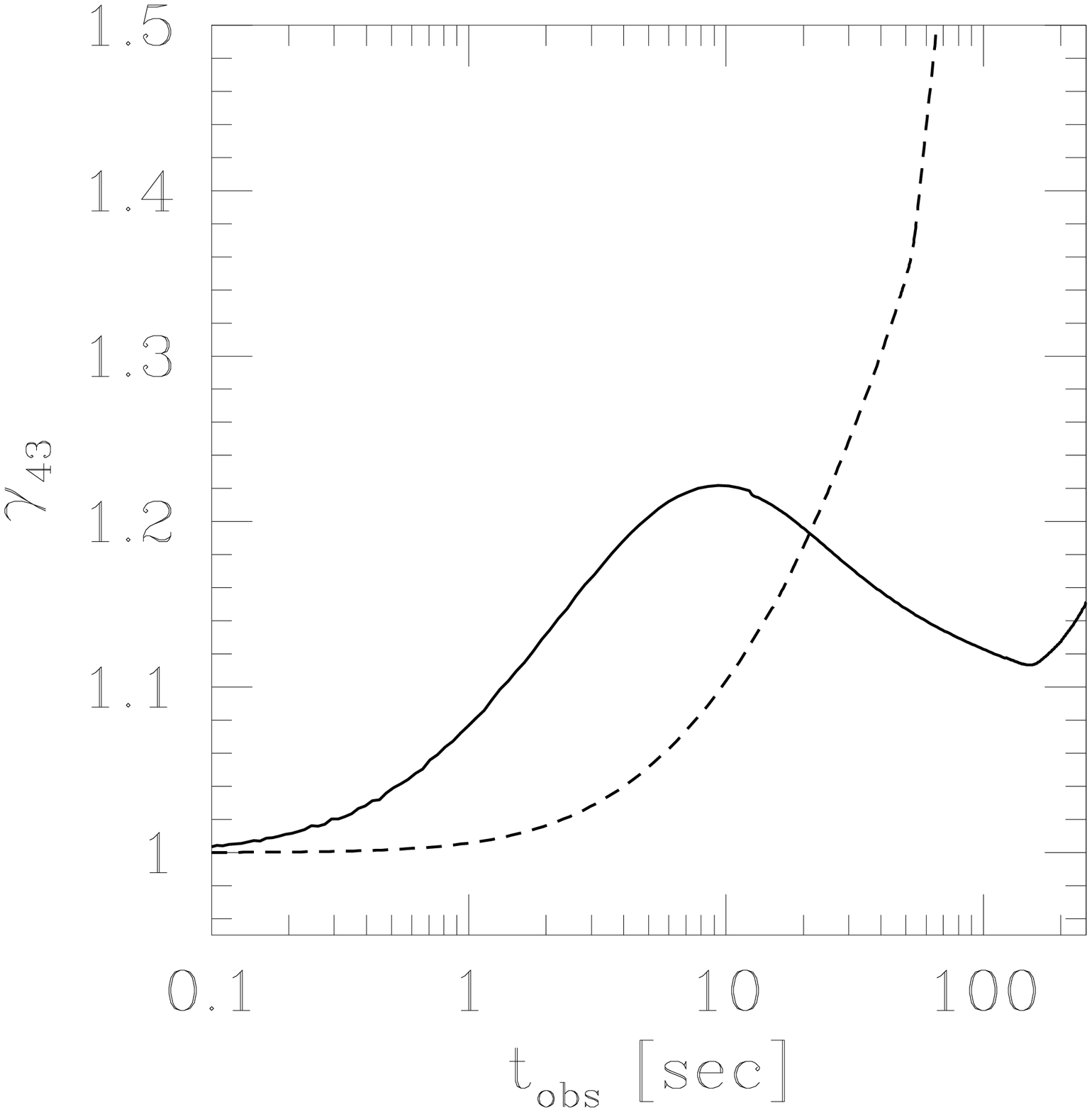, height=12pc, width=15pc} \end{array}$ 
   \caption{
{\bf Left}: Dashed curve shows the total energy of the blast wave 
calculated using $P_F=P_R$ assumption, solid curve -- using 
our mechanical model. 
{\bf Right}: Lorentz factor $\gamma_{43}$ of region 4 relative to region 3 
for the stratified (solid curve) and non-stratified (dashed curve)
fireballs; our mechanical model is used in the calculations.
}      
\end{figure} 


\subsection {3. Light curves from stratified fireballs} 

We calculate the afterglow emission using the standard 
simple model of the postshock plasma. It assumes that a fraction 
$\epsilon_e\sim 0.1$ of the shock energy goes to the electrons and 
accelerates them with a power-law energy spectrum $dN/dE_e\propto E_e^{-p}$.
The postshock magnetic field is parameterized by the ``equipartition''
parameter $\epsilon_B=(B^2/8\pi)U_{\rm th}^{-1}<1$ where  $U_{\rm th}$
is the thermal energy density of the shocked medium.

We discretize the external medium and the ejecta into spherical mass 
shells $m_i$ and use the Lagrangian description of the blast wave.
Each $m_i$ is impulsively heated at some point by a shock front
(forward or reverse). We track the subsequent evolution of the shell
and calculate its synchrotron emission in the shell comoving frame.

The shock-heated shell suffers radiative and adiabatic energy losses
in the expanding blast wave, and its magnetic field decreases, so
the shell emission weakens with time. We find its contribution to
observed radiation at a given observed time $t_{obs}$.
Integration over all $m_i$ gives the exact total spectrum
received from the blast wave at a given $t_{obs}$.

 
\begin{figure}[h] \centering $\begin{array}{cc} 
   \epsfig{file=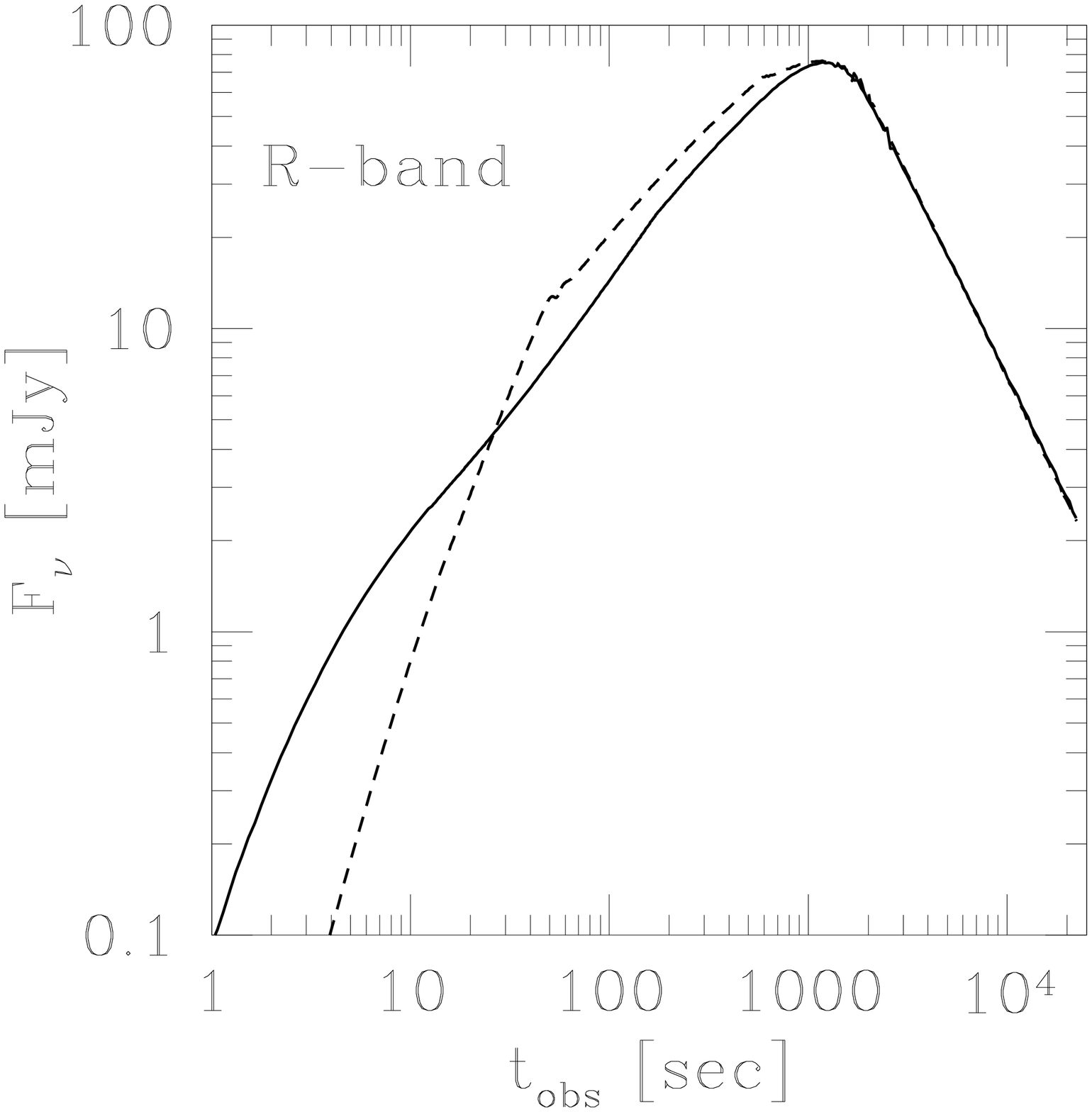, height=12pc, width=15pc}~ &
   \epsfig{file=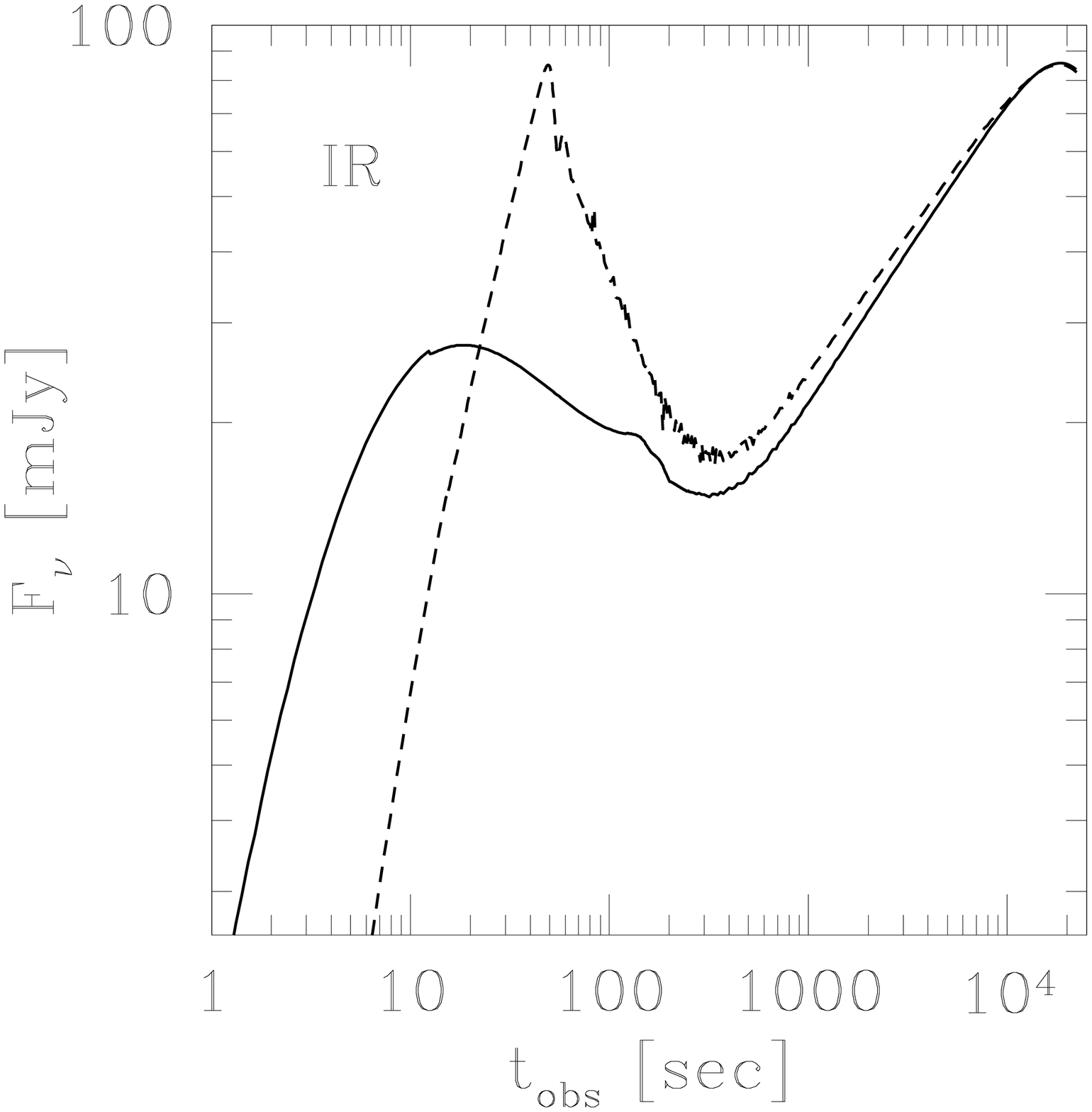, height=12pc, width=15pc} \end{array}$ 
   \caption{
{\bf Left}: Optical light curve (R-band)
from stratified (solid) and non-stratified (dashed) fireballs.
The explosion parameters are $\epsilon_B = 10^{-3}$, 
$\epsilon_e = 0.1$, $p=2.5$, $n_1 = 10$~cm$^{-3}$, 
and redshift $z=1$.  {\bf Right}: IR light curve ($\nu_{obs} = 10^{13}$~Hz)
from the same models.
}      
\end{figure} 

The optical and infrared light curves for two example models (stratified
and non-stratified) are shown in Figure 4. 
We find weak optical emission from the reverse shock.
It is weak because the RS Lorentz factor relative to the blast, 
$\gamma_{43}$, is modest and the emitted spectrum peaks in IR
rather than in optical band. 
The peak frequency is 
$$
 \nu_m\approx 10^6 B\gamma_m^2 {\rm ~Hz}, 
 \qquad
 \gamma_m=1+\frac{p-2}{p-1}\,\frac{m_p}{m_e}\,\epsilon_e\,(\gamma_{43}-1).
$$
The evolution of $\gamma_{43}$ is shown in Figure~3 (right panel); 
its typical value
is $1.2$ and the factor $(\gamma_{43}-1)^2$ that enters $\nu_m$ is
small, $\sim 1/20$, shifting the peak frequency to IR.

The behavior of the RS Lorentz factor $\gamma_{43}$ in the stratified
fireball is significantly different from the non-stratified case:
it reached maximum and then minimum (Fig. 3).
Therefore, the RS emission in the stratified fireball is weaker and its peak 
in the light curve is less pronounced, with a smoother rise and
slower decay.

\begin{theacknowledgments}
This work was supported by NASA grant NAG5-13382.
\end{theacknowledgments}

\end{document}